\documentclass[final,5p,times,twocolumn]{elsarticle}
\usepackage{amsmath}
\usepackage{xcolor}
\usepackage{accents}
\graphicspath{{figures/}}

\usepackage[sharp]{easylist}
\usepackage[switch]{lineno}
\usepackage[math]{cellspace}
\usepackage{mathtools}
\usepackage{inst_paper}
\usepackage{hyperref}
\journal{Classical and Quantum Gravity}

\begin{document}
\begin{frontmatter}

\title{Interferometric Constraints on Quantum Geometrical Shear Noise Correlations}

\author[fnal]{Aaron Chou}
\ead{achou@fnal.gov}

\author[fnal]{Henry Glass}
\ead{glass@fnal.gov}

\author[um]{H. Richard Gustafson}
\ead{gustafso@umich.edu}

\author[fnal,uc]{Craig J. Hogan}
\ead{cjhogan@fnal.gov}

\author[cal]{Brittany L. Kamai}
\ead{bkamai@caltech.edu}

\author[uc,kaist]{Ohkyung Kwon}
\ead{o.kwon@kaist.ac.kr}

\author[mit]{Robert Lanza}
\ead{rklanza@mit.edu}

\author[mit]{Lee McCuller}
\ead{mcculler@mit.edu}

\author[uc]{Stephan S. Meyer\corref{cor1}}
\ead{meyer@uchicago.edu}

\author[uc,um]{Jonathan W. Richardson}
\ead{jonathan.richardson@uchicago.edu}

\author[fnal]{Chris Stoughton}
\ead{stoughto@fnal.gov}

\author[fnal]{Ray Tomlin}
\ead{tomlin@fnal.gov}

\author[mit]{Rainer Weiss}
\ead{weiss@ligo.mit.edu}

\cortext[cor1]{Corresponding Author}
\address[fnal]{Fermi National Accelerator Laboratory}
\address[um]{University of Michigan}
\address[uc]{University of Chicago}
\address[cal]{California Institute of Technology}
\address[kaist]{Korea Advanced Institute of Science and Technology}
\address[mit]{Massachusetts Institute of Technology}

\begin{abstract}

Final measurements and analysis are reported from the first-generation Holometer, the first instrument capable of measuring correlated variations in space-time position at strain noise power spectral densities smaller than a Planck time. The apparatus consists of two co-located, but independent and isolated, 40~m power-recycled Michelson interferometers, whose outputs are cross-correlated to 25 MHz. The data are sensitive to correlations of differential position across the apparatus over a broad band of frequencies up to and exceeding the inverse light crossing time, 7.6~MHz. By measuring with Planck precision the correlation of position variations at spacelike separations, the Holometer searches for faint, irreducible correlated position noise backgrounds predicted by some models of quantum space-time geometry. The first-generation optical layout is sensitive to quantum geometrical noise correlations with shear symmetry---those that can be interpreted as a fundamental noncommutativity of space-time position in orthogonal directions. General experimental constraints are placed on parameters of a set of models of spatial shear noise correlations, with a sensitivity that exceeds the Planck-scale holographic information bound on position states by a large factor. This result significantly extends the upper limits placed on models of directional noncommutativity by currently operating gravitational wave observatories.

\end{abstract}

\end{frontmatter}

\twocolumn

\section{Introduction}
    \label{sec:intro}

It is often conjectured that classical space-time can be attributed to a statistical or emergent behavior of a fundamental quantum system \cite{Rovelli2004, Thiemann:2007zz, Ashtekar:2012np}. Although its degrees of freedom and dynamics are not known, theoretical studies of  information and  thermodynamics in black hole space-times appear to suggest that geometrical quantum states, and hence correlations,  are not localized in space and time, but can extend over macroscopic scales \cite{Jacobson1995,Verlinde2011,Padmanabhan:2013nxa,Jacobson:2015hqa}. If that is is also true for  states of nearly flat space-time, and if their correlations affect the propagation of light,  it may be possible to measure exotic quantum correlations  of  geometry in a macroscopic laboratory apparatus of sufficient sensitivity, in much the same way that ``spooky'' spacelike correlations of particle properties have been  studied on macroscopic scales. 

 In this paper, we report measurements of correlations of variations in the space-time positions of massive bodies at an unprecedented level of precision.  Because the precision of the correlation measurement  corresponds to a quantity of information that is much higher than the gravitational value (one quarter of the bounding area in Planck units), it  can be interpreted as a deep probe of symmetries of quantum degrees of freedom not of particles, but of space-time itself.

The magnitude of such quantum geometrical noise correlations can be estimated by straightforward extrapolations of quantum mechanics and relativity \cite{Hogan:2008a, Hogan:2008b,  Hogan:2012}. A natural measure is the dimensionless noise power spectral density of fractional position displacements:
\begin{align}
\label{eq:powersd}
h^2(f, t) &\equiv \int_{-\infty}^{\infty} \left\langle \frac{\delta\hspace{-.1em}L(t)}{L} \, \frac{\delta\hspace{-.1em}L(t - \tau)}{L} \right\rangle e^{- 2\pi i \tau f} \, d\tau \;.
\end{align}
Here, fractional position fluctuations $\delta L/L$, known as strain, refer to displacements of objects (such as mirrors) separated from a center object by distance $L$\footnote{In Eq.~\ref{eq:powersd}, brackets indicate the expectation over the distribution of the random signal $\delta\hspace{-.1em}L(t)$, which need not be stationary. For a stationary signal, Eq.~\ref{eq:powersd} reduces to an explicitly time-independent form, $h^2(f,t)=h^2(f)$.}. In these units, the expected scale of exotic correlations, normalized by the information in black hole event horizons on scale $L$, is about a Planck time, $t_P=\sqrt{\hbar G/c}= 5.4\times 10^{-44}$ sec.  This information density can be achieved by displacements of order a Planck length that are  uncorrelated at timelike separation separation greater than $t_P$, but  have spacelike correlations on the scale $L$ of an apparatus.
The accumulated measurable physical displacement over a light crossing time approximates a random walk of about Planck length per Planck time, which can  be much larger than a Planck length, and hence accessible to experimental study.
 
The only experimental technique currently able to measure displacements with this precision is laser interferometry \cite{Adhikari:2014}. In this case, the spectrum $h^2$ of an interferometer signal measures the displacement of mirrors in some arrangement in space. The layout of the light paths determines the coupling of correlated geometrical fluctuations to the optical phase. In general, that mapping cannot be described by standard physics based on a classical metric, but testable predictions for signal correlations have been made using models based on general principles of symmetry and causality \cite{Kwon:2014, Hogan:2015a, Hogan:2015b, Hogan:2016}. Thus there is an opportunity for experiments to measure or constrain possible quantum departures from classical space-time structure, by measuring with Planck precision the coherence of spacelike-separated geodesics of mirrors.

The Holometer \cite{Holo:Instrument} is the first instrument built to enable experimental studies of quantum noise correlations in the positions of bodies (mirrors) in space-time, at strain noise power spectral densities smaller than a Planck time. It consists of two co-located, but independent and isolated, 40~m power-recycled Michelson interferometers, whose outputs are cross-correlated to 25~MHz. The data are sensitive to correlations of differential position across the apparatus over a broad band of frequencies up to and exceeding the inverse light crossing time, 7.6~MHz. Correlations at both spacelike and timelike separations, determined by the optical layout of the two interferometers, are measured in the time dimension represented by the signal streams.

While based largely on technology developed for gravitational wave detectors, the Holometer operates at much higher sampling frequency and bandwidth (MHz compared to kHz). Existing gravitational wave detectors, such as LIGO, VIRGO, and GEO600, constrain some hypotheses about quantum geometrical decoherence, particularly those based on fluctuations of the metric \citep{Kwon:2014}. However, since those detectors are specifically optimized to measure a metric strain at frequencies small compared their inverse light travel time (timelike separations), a much larger space of possible quantum geometrical noise arising from Planck-scale decoherence remains unexplored.

This paper presents the final results of the first-generation Holometer search for quantum geometrical noise correlations with shear symmetry---those that can be interpreted as a fundamental noncommutativity of space-time position in orthogonal directions. This work extends the early results reported in \cite{Holo:PRL} with a factor of five more data, which now constrain a broader class of spatial shear noise correlation models. With $2\sigma$ significance, the final data probe nearly an order of magnitude below the level of correlated position noise predicted for a Planck-scale, shear-symmetry holographic information bound on space-time position states, with no evidence of a correlation detected. Stringent experimental constraints are placed on parameters of a set of models of universal spatial shear noise correlations. This result significantly extends the upper limits on models of directional noncommutativity from currently operating gravitational wave observatories.

\section{Experimental Design}

Unlike gravitational wave detectors, the outputs of the Holometer interferometers are sampled faster than the light crossing time, $L/c \approx 130$~ns, to resolve the causal time scale of the $L=40$~m system. This scale is of special significance because, in causality-preserving models of Planck-scale positional decoherence \cite{Hogan:2008a, Hogan:2008b,  Hogan:2012, Kwon:2014, Hogan:2015a, Hogan:2015b, Hogan:2016}, it imposes a minimum transition time on the geometrical state of the extended system. This sets the time scale of fluctuations in the apparent position of the distant end mirrors relative to the center optic. Thus interferometer measurements separated in time or space by less than $L/c$ or $L$, respectively, are predicted to exhibit highly correlated fluctuations. The Holometer directly searches for these correlations by cross-correlating the outputs of the two closely neighboring instruments\footnote{In the event of a detection, translationally separating the two interferometers by a distance $>L$ is predicted to then null the measured correlation, providing an experimental control.}. A more detailed discussion of the Holometer instrumentation, including systematic error analysis and characterization of environmental backgrounds, is presented separately in \cite{Holo:Instrument}. The data presented herein include the systematic error estimates of \cite{Holo:Instrument}, typically 10-15\% per frequency bin, which are added in quadrature to the statistical error.

\begin{figure}[!tp]
	\centering
	\includegraphics[width=1\linewidth]{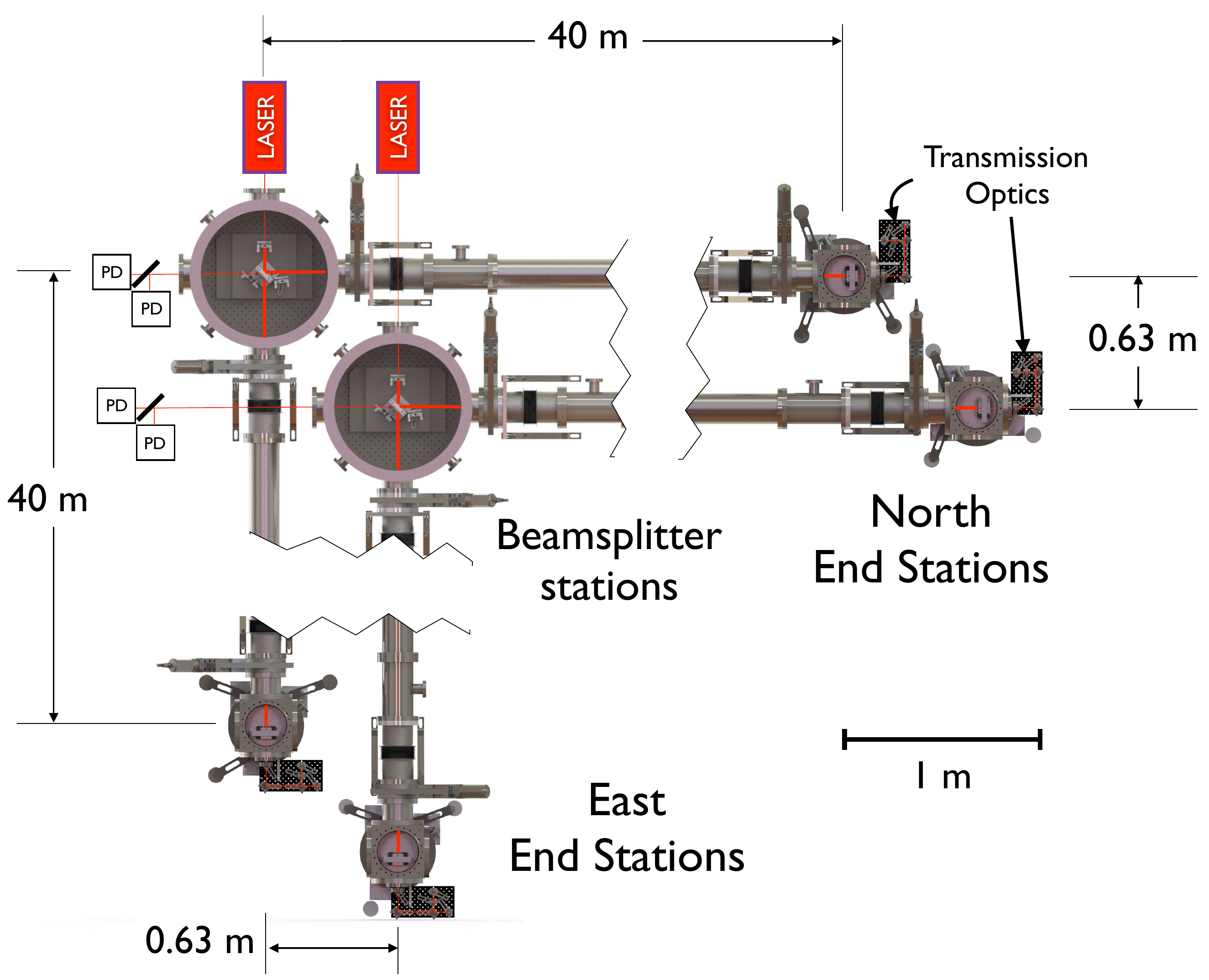}
	\caption[Spatial orientation of the two independent 40~m interferometer systems.]{Spatial orientation of the two independent 40~m interferometer systems. Each interferometer has straight, orthogonally-oriented arms of length 40~m, with a translational separation of 0.90~m between the two beamsplitters. In each system, a $\lambda=1064$~nm laser beam, represented by the red line, is injected onto the beamsplitter and traverses two orthogonal 80~m roundtrip paths simultaneously. Small fluctuations in the relative distance from the beamsplitter to the $x$- and $y$-mirrors ($\delta L << \lambda$) linearly modulate the power of the recombined output beam, which is split and directed onto two photoreceivers. The cross-correlation of the high-frequency photoreceiver signals is sensitive to quantum geometrical noise correlations with shear symmetry---those that can be interpreted as a fundamental noncommutativity of space-time position in two orthogonal directions.}
	\label{fig:layout}
\end{figure}

In its first-generation layout, the Holometer consists of two identical, but independent and isolated, Michelson interferometers with straight, orthogonally-oriented 40~m arms. The two interferometers are co-aligned in space at a translational separation of 0.90~m. Figure~\ref{fig:layout} shows this configuration. A 1~W, $\lambda=1064$~nm laser beam is injected onto a beamsplitter and traverses two orthogonal 80~m roundtrip paths simultaneously. Small fluctuations in the relative distance from the beamsplitter to the $x$- and $y$-mirrors ($\delta L << \lambda$) linearly modulate the power of the recombined beam. To improve quantum-limited sensitivity to differential position displacements, a power-recycling mirror is placed at the symmetric (input) port of each interferometer to form an effective Fabry-Perot cavity with its end mirrors, generating a 2~kW storage beam incident on the beamsplitter. The interferometers are operated at a differential arm length offset of 1~nm chosen to output a 200~mW beam at the antisymmetric (output) port, which is split and directed onto two photoreceivers modified to each absorb 100~mW of incident power. At this operating point, the interferometers output approximately equal parts signal-carrying interference power and defect leakage power.

A high-throughput data processing system samples the four antisymmetric-port readout signals, along with four environmental monitoring channels, at 50~MHz and computes the cross-spectral density (CSD) matrix between all eight channels. The cross-spectra are continuously averaged over time to achieve a reduction in measurement variance proportional to the total integration time. The spectra of signals in the basis of the two interferometers are then reconstructed from a variance-weighted average of the spectra measured in the basis of the four photodetector signals. Quantum geometrical noise will appear as an irreducible correlated noise floor in the interferometer cross spectrum.

\section{Data}
\label{sec:data} 

\begin{figure}[!tp]
	\centering
	\includegraphics[width=1\linewidth]{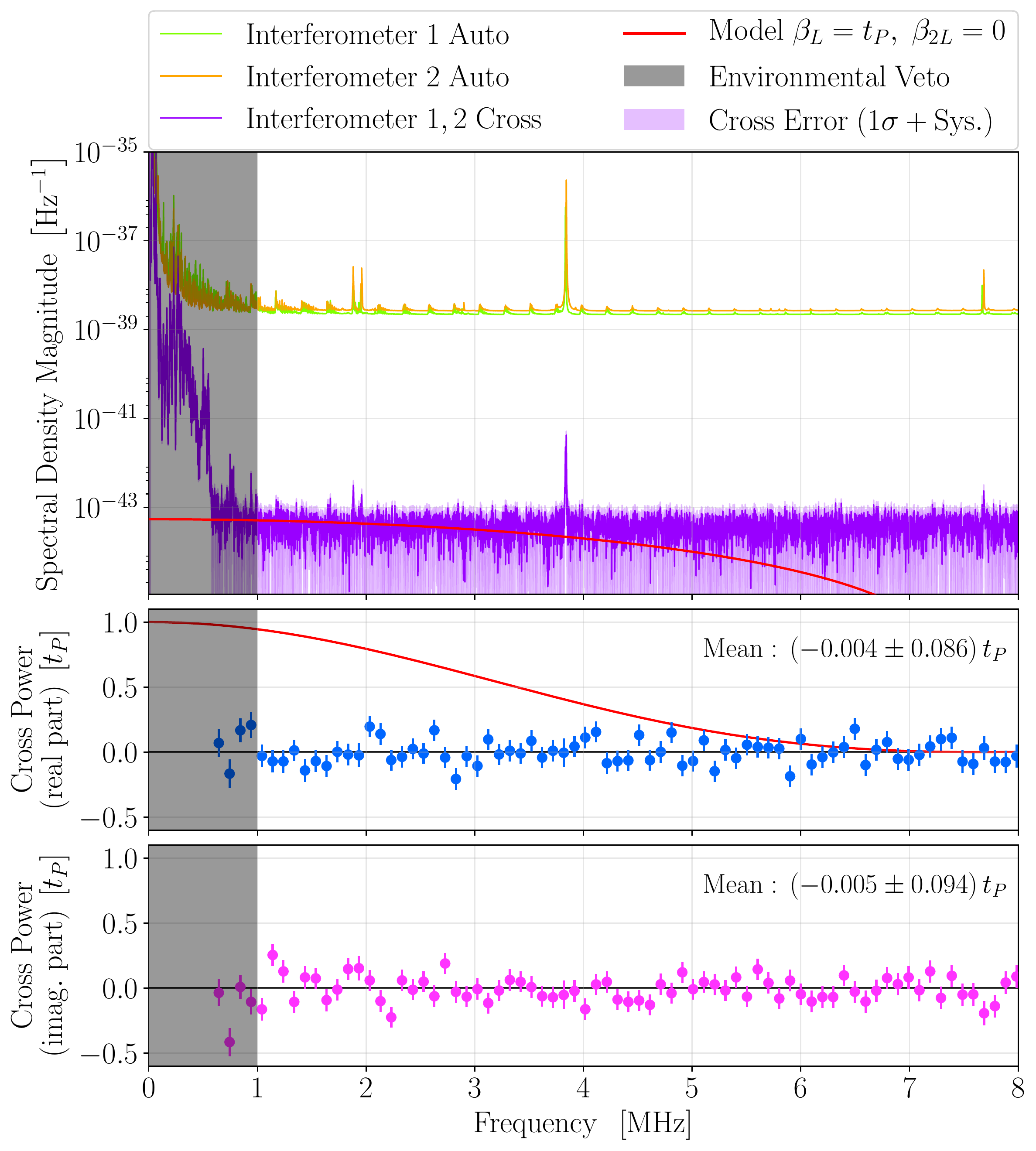}
	\caption[Cross-spectral density measurement of the signals of two isolated, co-located 40~m Michelson interferometers with straight arms, averaged over an integration time of 704~hours.]
{
Cross-spectral density measurement of the signals of two isolated, co-located 40~m Michelson interferometers with straight arms, after an integration time of 704 hours. Top panel: The magnitude of the auto and cross interferometer signal spectra at high frequency resolution (1.9~kHz). Bottom panels: The interferometer cross-spectrum at lowered frequency resolution (100~kHz), separated into its real (middle panel) and imaginary (bottommost panel) parts. In all panels, the uncertainties denoted by shaded regions or error bars refer to the combined $1\sigma$ statistical error and systematic error, with the errors added in quadrature. Both real and imaginary parts of the cross spectrum measure an average in-band noise power consistent with zero.
}
	\label{fig:data}
\end{figure}

Figure~\ref{fig:data} shows the measured signal spectra of the two isolated, co-located interferometers after an integration time of 704~hours. These data were taken between July 2015 and February 2016. The initial 145~hours of data, first reported in \cite{Holo:PRL}, are included again here together with an additional 560~hours of new data. The signal spectra in the top and bottom two panels are shown at resolutions of 1.9~kHz and 100~kHz, respectively, rebinned from their original frequency bin width of 380~Hz using an estimator which accounts for the bin-bin covariance due to spectral leakage. Every data point is thus a statistically independent measurement, which will be seen to simplify the inference of model constraints in \S\ref{sec:analysis}.

The top panel shows the magnitude of the interferometer signal spectra at high frequency resolution (1.9~kHz). The auto-spectra (green and orange curves) measure the total differential position noise power in each interferometer from all sources, while the cross-spectrum (purple curve) measures only the contribution from noise sources common to both instruments. The shaded region enclosing each curve denotes the combined $1\sigma$ statistical error and systematic error from \cite{Holo:Instrument}, with the errors added in quadrature. The strong correlations below 1~MHz are due to environmental background leakage into the amplitude and phase noise of the lasers (see \cite{Holo:Instrument}). This band is vetoed from analysis, as indicated by the shading. Above 1~MHz, optical shot noise is the dominant source of noise power at nearly every frequency, and the spectra are well-described by Gaussian noise. A repeating sequence of peaks is due to thermally excited acoustic modes of individual optics substrates and mode resonances of the Fabry-Perot cavity. Because the interferometers use independent optics and lasers, the excess noise from these sources is uncorrelated, but it does reduce the experimental sensitivity at affected frequencies.

The bottom two panels show the interferometer cross-spectrum at lowered frequency resolution (100~kHz), separated into its real (middle panel) and imaginary (bottommost panel) parts. As in the top panel, error bars denote the combined $1\sigma$ statistical error and systematic error, with the errors added in quadrature. Quantum geometrical shear noise would appear as an irreducible correlated noise floor in the real part of the cross spectrum at a level near the Planck time. The noise spectrum is expected to be purely real because common geometrical fluctuations will be detected in-phase by co-located instruments. An example shear noise model spectrum is shown by the red curve for the spectral parameters $\beta_L=t_P$ and $\beta_{2L}=0$ (see Eq.~\ref{eq:generalized_model}). The imaginary part of the cross-spectrum provides a pure measure of the statistical uncertainty. Rather than converging on a non-zero value, the cross-spectrum is statistics-limited above 1~MHz, with Planck spectral sensitivity ($h^2 < t_P$) across a broad band. In the following section, these data are used to place stringent constraints on a general model of spatial shear noise correlations.

\section{Model Constraints}
\label{sec:analysis}

The interferometer cross-spectrum data constrain models of quantum geometrical noise correlations respecting shear symmetry, or any model which can be interpreted as a fundamental noncommutativity of space-time position in orthogonal directions. In a review of proposed models, \cite{Hogan:2015a} notes that, regardless of detailed dynamics, the interferometer response can depend, at most, on only two time scales: the light crossing time, $L/c$, and the round-trip light crossing time, $2L/c$. Accordingly, for experimental tests \cite{Hogan:2015a} proposes a generalized two-parameter model of spatial shear noise correlations, whose spectrum is the linear superposition of two sinc-square terms governed by these canonical time scales.

This paper considers a slightly modified form of this spectral model,
\begin{align}
\label{eq:generalized_model}
h^2\left(f;\beta_{L},\beta_{2L}\right) &= \beta_{L} \, {\rm sinc}^2\left( \frac{f}{c/\pi L} \right) + \beta_{2L} \, {\rm sinc}^2\left( \frac{f}{c/2\pi L} \right) \;,
\end{align}
which has been converted to dimensionless units of strain and its normalization simplified. The normalization parameters, $\beta_{L}$ and $\beta_{2L}$, are predicted to obey $\beta_{L}+\frac{1}{2} \beta_{2L} \approx t_P$ under a Planck-scale, shear-symmetry holographic information bound on space-time position states. Requiring the superposed spectrum to result from a stationary process acting on a real-valued time series (the antisymmetric-port power) imposes the additional physicality constraints\footnote{These constraints follow from requiring the total power spectral density $h^2\ge0$ for all frequencies.} $\beta_{L}\ge 0$ and $\beta_{2L} \ge -\beta_{L}$.

A Bayesian inference of the measured interferometer cross-spectrum is performed to obtain constraints on this generalized model. Adopting a flat prior, the joint posterior of the two-parameter spectral model (Eq.~\ref{eq:generalized_model}) is given by
\begin{align}
P\left(\beta_{L},\beta_{2L} \,\big| \,\mathcal{R}\left[{\rm CSD}\right]\right) &= \frac{\displaystyle {\rm exp}\left(-\,\frac{\chi^2\left(\beta_{L},\beta_{2L}\right)}{2}\right)}{\displaystyle \int {\rm exp}\left(-\,\frac{\chi^2\left(\beta_{L},\beta_{2L}\right)}{2}\right) \,d\beta_{L} \,d\beta_{2L}}
\end{align}
with $\chi^2$ statistic
\begin{align}
\chi^2\left(\beta_{L},\beta_{2L}\right) &= \sum_{i=1}^{N} \frac{\left(\mathcal{R}\left[{\rm CSD}(f_i)\right] - h^2\left(f_i;\beta_{L},\beta_{2L}\right)\right)^2}{{\rm Var}\,\Big(\mathcal{R}\left[{\rm CSD}(f_i)\right]\Big)} \;.
\end{align}
As denoted above, the model constraints depend only on the real part, $\mathcal{R}\left[{\rm CSD}\right]$, of the complex cross-spectrum. This reflects the fact that geometrical fluctuations of the region of space-time commonly occupied by the two interferometers are detected in-phase.

\begin{figure}[!tp]
	\centering
	\includegraphics[width=.8\linewidth]{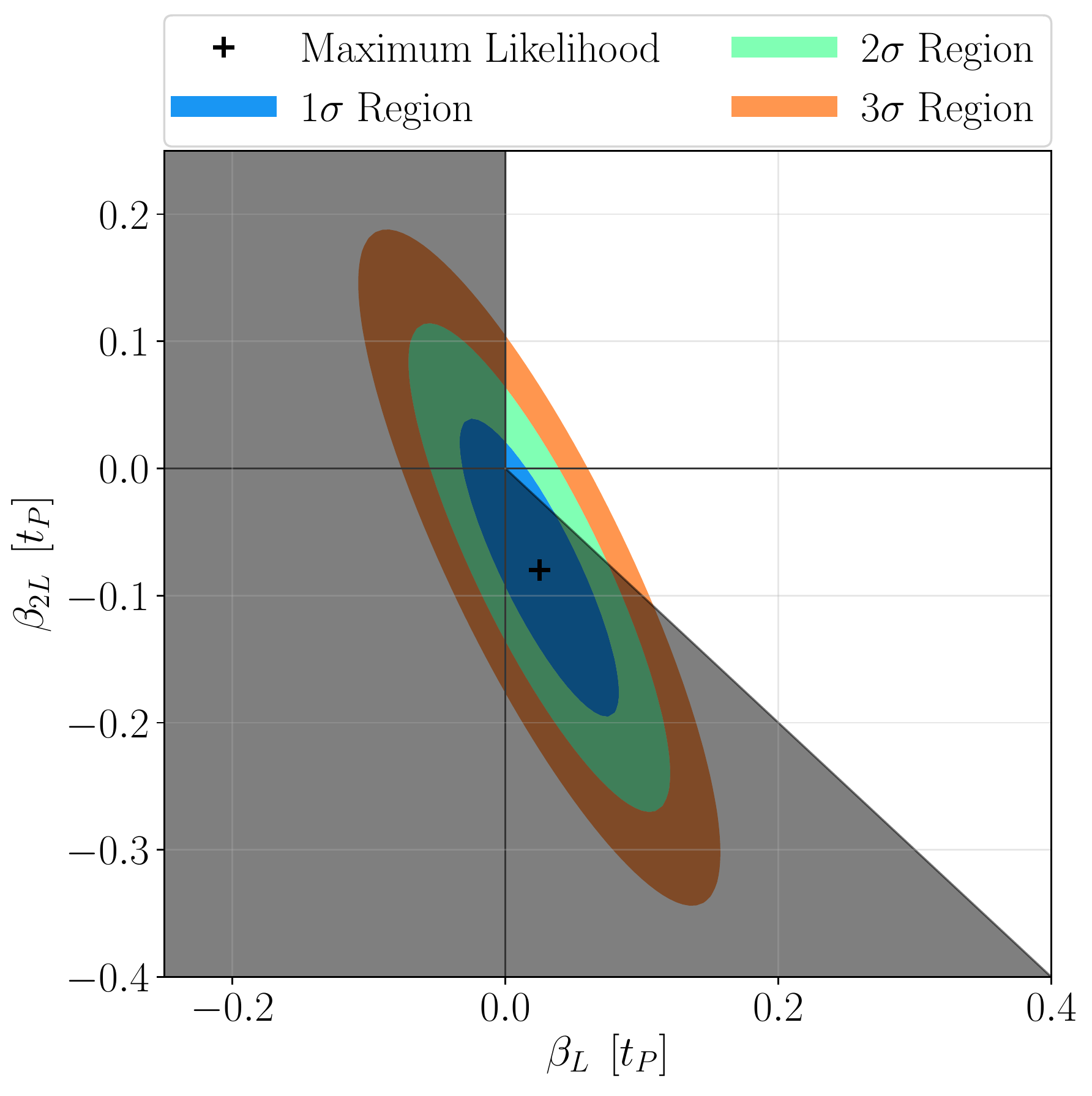}
	\caption[Constraints on a two-parameter spectral model of universal spatial shear noise correlations.]{Constraints on a two-parameter spectral model of universal spatial shear noise correlations. Each model spectrum is a linear superposition of two sinc-square terms with characteristic time scales $L/c$ and $2L/c$, whose relative amplitudes are specified by the parameters $\beta_{L}$ and $\beta_{2L}$, respectively. The greyed area is excluded on physicality grounds as discussed in \S\ref{sec:analysis}. The Holometer data are consistent at $1\sigma$ with the classical space-time model, $\beta_{L}=\beta_{2L}=0$. At $2\sigma$ significance, the data limit the amplitudes of both spectral correlation terms, $\beta_{L}$ and $\beta_{2L}$, to well below the predicted scale of quantum geometrical position noise, which is of order unity.}
	\label{fig:constraints}
\end{figure}

Figure~\ref{fig:constraints} shows the joint posterior obtained from modeling the interferometer cross-spectrum measurements. The $1\sigma$, $2\sigma$, and $3\sigma$ confidence intervals are denoted by the blue, green, and red contours, respectively. The greyed area denotes unphysical models, those whose spectra cannot be the response to a stationary (time-invariant) statistical process acting on the antisymmetric-port power. The zero-amplitude model, $\beta_{L}=\beta_{2L}=0$, corresponds to classical space-time structure, under which no geometrical fluctuations occur. The Holometer data are consistent at $1\sigma$ with the classical space-time model. At $2\sigma$ significance, the data limit the amplitudes of the spectral correlation terms to $\left|\beta_{L}\right| < 0.10 \, t_P$ and $\left| \beta_{2L} \right| < 0.25 \, t_P$, well below the predicted scale of quantum geometrical position noise. The difference in upper limits arises from the bandwidth difference of the two sinc-square terms, 7.6~MHz for $\beta_{L}$ versus 3.8~MHz for $\beta_{2L}$.

\section{Conclusions}

The first-generation Holometer has tested and conclusively excluded a general class of models of quantum geometrical shear noise correlations. Geometrical noise of this symmetry can be interpreted as a fundamental noncommutativity of positions in space and time. This noise would appear as an irreducible correlated noise in the cross spectrum of two co-located interferometers, with a strain noise power spectral density of order the Planck time. The cross-spectrum measurement is statistics-limited above 1~MHz and constrains such noise to well below Planck spectral density ($h^2 < t_P$) across a broad band, with no evidence of a correlation detected. 

A new Lorentz invariant model for exotic correlations \cite{Hogan:2015b, Hogan:2016}, based on a classical causal structure around any observer, predicts a  different symmetry from the model tested here. According to this hypothesis, the structure of exotic spacelike correlations is purely {rotational} about the world line of any observer. The first-generation Holometer, by the design of its optical geometry, has no sensitivity to such rotational effects, so the result reported here does not constrain this model. A reconfigured, second-generation apparatus with sensitivity to rotational forms of geometrical noise is currently under construction at Fermilab. It is anticipated to become operational in mid 2017.

\section*{Acknowledgements}

This work was supported by the Department of Energy at Fermilab under Contract No. DE-AC02-07CH11359 and the Early Career Research Program (FNAL FWP 11-03), and by grants from the John Templeton Foundation, the National Science Foundation (Grants No. PHY-1205254, No. DGE-0909667, No. DGE-0638477, and No. DGE-1144082), NASA (Grant No. NNX09AR38G), the Fermi Research Alliance, the Ford Foundation, the Kavli Institute for Cosmological Physics, University of Chicago/Fermilab Strategic Collaborative Initiatives, and the Universities Research Association Visiting Scholars Program. O.K. was supported by the Basic Science Research Program (Grant No. NRF-2016R1D1A1B03934333) of the National Research Foundation of Korea (NRF) funded by the Ministry of Education. The Holometer team gratefully acknowledges the extensive support and contributions of Bradford Boonstra, Benjamin Brubaker, Marcin Burdzy, Herman Cease, Tim Cunneen, Steve Dixon, Bill Dymond, Valera Frolov, Jose Gallegos, Emily Griffith, Hartmut Grote, Gaston Gutierrez, Evan Hall, Sten Hansen, Young-Kee Kim, Mark Kozlovsky, Dan Lambert, Scott McCormick, Erik Ramberg, George Ressinger, Doug Rudd, Geoffrey Schmit, Alex Sippel, Jason Steffen, Sali Sylejmani, David Tanner, Jim Volk, William Wester, and James Williams for the design and construction of the apparatus.

\section*{References}

\bibliographystyle{elsarticle-num}
\bibliography{holometer}

\end{document}